\documentclass[11pt,preprint]{aastex}

\def\gtrsim{\mathrel{\hbox{\rlap{\hbox{\lower4pt\hbox{$\sim$}}}\hbox{$>$}}}}
\def\lesssim{\mathrel{\hbox{\rlap{\hbox{\lower4pt\hbox{$\sim$}}}\hbox{$<$}}}}
\def\gtrsim{\mathrel{\hbox{\rlap{\hbox{\lower4pt\hbox{$\sim$}}}\hbox{$>$}}}}
\def\lesssim{\mathrel{\hbox{\rlap{\hbox{\lower4pt\hbox{$\sim$}}}\hbox{$<$}}}}

\slugcomment{To appear in {\it The Astrophysical Journal, Letters}}

\shortauthors{D. C. Homan}
\shorttitle{Inverse Depolarization}
\begin{document}
\title{Inverse Depolarization: A Potential Probe of Internal Faraday Rotation 
and Helical Magnetic Fields in Extragalactic Radio Jets}

\author{D. C. Homan}

\affil{Department of Physics and Astronomy, Denison University,
Granville, OH 43023; homand@denison.edu}

\begin{abstract}
Motivated by recent observations that show increasing fractional linear 
polarization with increasing wavelength in a small number of optically thin jet 
features, i.e. ``inverse depolarization'', we present a physical model that 
can explain this effect and may provide a new and complementary probe of the 
low energy particle population and possible helical magnetic fields in 
extragalactic radio jets. In our model, structural inhomogeneities in 
the jet magnetic field create cancellation of polarization along the line of 
sight.  Internal Faraday rotation, which increases like wavelength squared, 
acts to align the polarization from the far and near sides of the jet, leading 
to increased polarization at longer wavelengths.  Structural inhomogeneities of 
the right type are naturally produced in helical magnetic fields and will also 
appear in randomly tangled magnetic fields. We explore both alternatives and find 
that, for random fields, the length scale for tangling cannot be too small a 
fraction of the jet diameter and still be consistent with the relatively high 
levels of fractional polarization observed in these features. We also find that 
helical magnetic fields naturally produce transverse structure for inverse 
depolarization which may be observable even in partially resolved jets. 
\end{abstract}


\section{Introduction}

Faraday rotation, the rotation of the plane of linearly polarized
radiation propagating through a magnetized plasma, has
long been thought not to play an important role {\em internal}
to extragalactic radio jets, in large part because the
depolarization and divergence from a pure $\lambda^2$ rotation 
law expected due to large amounts of internal rotation \citep{B66} 
have not been clearly observed; however, detecting this behavior
in real sources can be difficult \citep{CJ80}.
Observations of circularly polarized
radiation have suggested that Faraday rotation may indeed
operate internal to jets to drive the conversion of linear
polarization into circular \citep[e.g.][]{WH03,HL09}. 
Additionally, observations of variations in Faraday 
rotation and transverse structure, or gradients, in rotation
measure have indicated that some Faraday screens 
lie very close to the jet, perhaps in a boundary layer
around it that may contain helical magnetic fields \citep[e.g.][]{A02,
G04,ZT05,G11,K11}.  The observation and interpretation of
transverse rotation measure gradients as evidence for helical fields
remains controversial \citep{TZ10,BM10}. Given the central role 
helical fields are believed to play in collimating and accelerating 
jets near their
origin from the super massive black hole--accretion disk 
system \citep[e.g.][]{M01}, it is important 
to pursue additional observational signatures of helical 
magnetic fields in jets \citep{LPG05,MJ08,CB11,P11}.

An intriguing phenomenon has recently been observed in 
multiple optically thin jet features in the blazars 3C\,273, 
3C\,454.3, and a single jet feature in the blazar 1514$-$241.
\citet{HL12} have measured increasing polarization with
increasing wavelength in their Very Long Baseline Array 
observations of these jet features at $\lambda\lambda$ 2.0, 2.5, 3.6, 
and 3.7 cm. The observed increases range from factors 
of $\sim 1.5$ up to $\sim 3$ times over this wavelength
range.  \citet{H02} saw a similar increase in fractional polarization
of up to a factor of two in a jet feature in 3C\,120 between 
$\lambda\lambda$ 1.3 
and 2.0 cm. The jet features involved in these cases are relatively 
strongly polarized at their maximum wavelength, with fractional
polarization ranging from $5$ to more than $20$ percent.  The effect
can apparently vary with epochs separated by a few months, although it
is clearly observed for two epochs spanning three months in two 
features in 3C\,273 \citep{HL12} and over four epochs 
spanning six months in the jet feature in 3C\,120 \citep{H02}.

The sense of this effect is the opposite of that normally
expected for Faraday depolarization where fractional polarization
decreases with increasing wavelength, and we therefore term 
it ``inverse depolarization''.  Increasing fractional
polarization with wavelength might naturally arise in the unresolved, optically thick 
core region where one is observing different parts of the jet at different
wavelengths \citep[e.g.][]{K81}, but the jet features showing this 
effect are (1) well separated from the core in the optically 
thin part of the jet and (2) reasonably isolated from other 
strong jet features \citep{HL12}.  In this letter we 
propose an alternative explanation which combines structural
inhomogeneities in the jet magnetic field along the line of sight
with {\em internal} Faraday rotation to produce a inverse
depolarization effect in optically thin jet features\footnote{A similar 
effect due to Faraday conversion of linear to 
circular polarization was seen in simulations by \citet{BH09} at modest 
optical depths, $\tau\simeq 0.1-1.0$, while
the mechanism proposed here can operate in optically thin regions
and requires only internal Faraday rotation.}.  If our
explanation is correct, observed inverse depolarization is
evidence of internal Faraday rotation in jets and may prove to be 
a powerful and complementary tool to rotation measure studies 
of jet magnetic fields and particles, including helical 
field geometries.  

Our model for inverse depolarization is 
presented in \S{2} and applied to both helical field geometries 
and randomly tangled magnetic fields. \S{3} discusses these
results and the prospects for using this effect to study the magnetic
field geometry of extragalactic jets.

\section{Model for Inverse Depolarization}

In the subsections that follow, we present a simple model for 
generating inverse depolarization, i.e. increasing polarization
with increasing wavelength, in radio jets at low optical depth, 
$\tau \ll 1$.  Our model combines {\em internal} Faraday rotation
with structural differences in the magnetic field between the far
and near sides of the jet.  If the magnetic field structure 
naturally leads to low net polarization at short wavelengths, internal
Faraday rotation can act to align polarization from the far side 
of the jet with polarization produced at the near side 
of the jet, reducing the cancellation between them and leading to 
increased net polarization at longer wavelengths.

We characterize the amount of internal rotation at a given wavelength by
the Faraday depth, $\tau_f$, which is proportional to $\lambda^2$.  
Internal to the jet, Faraday depth may be due to either a population 
of thermal electrons or the low-energy end of the relativistic 
particle distribution \citep[e.g.][]{JOD77,HS11}, and we discuss
these contributions in \S{\ref{s:tau_f}}.  

\subsection{Linearly Polarized Transfer at Low Optical Depth}

The radiative transfer equations for Stokes $Q$ and $U$ at low 
optical depth are given by the following expressions, assuming the projected 
magnetic field is purely east-west \citep{JOD77,Jones88}. Note
that in the low optical depth limit, we have ignored the coupling of 
Stokes $I$ and $V$ to $Q$ and $U$

\begin{equation}
\frac{dQ}{d\tau} + Q + \zeta_{V}^*U=\epsilon J, \,\,\,\,
\frac{dU}{d\tau} + U - \zeta_{V}^*Q=0
\end{equation}

\noindent The solutions to these equations\footnote{The solutions 
to the general form were obtained with WolframAlpha, 
http://www.wolframalpha.com/ and checked against numerical
integration of the full equations of radiative transfer \citep{HL09}.}, 
including an initial polarization, $P_0$ at angle $\chi_0$ are

\begin{eqnarray}
Q = \frac{\epsilon J}{(\tau_f/\tau)^2+1}\left[1-e^{-\tau}\cos{\tau_f}+\left(\frac{\tau_f}{\tau}\right)e^{-\tau}\sin{\tau_f}\right]
+P_0e^{-\tau}\cos(2\chi_0+\tau_f)
\nonumber\\
\\
U = \frac{\epsilon J}{(\tau_f/\tau)^2+1}\left[\left(\frac{\tau_f}{\tau}\right)(1-e^{-\tau}\cos{\tau_f})-e^{-\tau}\sin{\tau_f}\right]
+P_0e^{-\tau}\sin(2\chi_0+\tau_f)
\nonumber
\end{eqnarray}

\noindent where $\tau$ is the optical depth and $\tau_f=\zeta_{V}^*\tau$ 
is the Faraday depth and is proportional to $\lambda^2$ \citep[e.g.][]{JOD77,CJ80}. Note 
that at low optical depth, Stokes $I = J\tau$.  The final Stokes 
$Q$ and $U$ are used to compute the fractional 
polarization, $m = \sqrt{Q^2+U^2}/I$, and electric vector position 
angle (EVPA), $\chi=0.5\times\tan^{-1}(U/Q)$ of the emergent
polarization.

\subsection{Inverse Depolarization in a Simple Two-Cell System}

We apply these expressions to a ``two-cell'' system along the line of sight.  
Each cell has an internal Faraday rotation of depth $\tau_f/2$ for a total 
Faraday depth of $\tau_f$ for the system.  In addition to their common
line-of-sight magnetic field, both cells also have a uniform 
magnetic field, $B_\perp$, in the plane of the sky.  The field in the two 
cells differs only by a rotation on the sky of $\Delta\phi$.  The $B_\perp$ 
in the cell closest to the observer is taken to be purely East-West, while 
the field in the furthest cell differs in orientation by an offset angle, $\Delta\phi$, 
Northward from East.  Note that positive Faraday depth will tend to rotate 
the $\chi$ of the emitted radiation from North to East\footnote{Negative Faraday
depth would simply result in a rotation of $\chi$ in the opposite direction.}.

Polarized radiation generated in the furthest cell must travel through the 
nearest cell to reach the observer.  In the absence of internal  
Faraday rotation, the offset angle $\Delta\phi$ will produce some cancellation 
between the polarization emitted from the far and near cells.  For 
$\Delta\phi = 90^\circ$, this structural cancellation would 
be complete at short wavelengths. 

Figure 1 shows the emergent polarization as a function of 
$\tau_f$ ($\propto \lambda^2$) for a range of $\Delta\phi$ offsets.  
For $\Delta\phi=0$, \citet{B66} depolarization in a slab geometry is reproduced.  
At larger offsets, partial cancellation of the emergent polarization is 
clear at $\tau_f = 0$, and for $0 < \Delta\phi \leq 90^\circ$ the internal rotation 
initially begins to reverse this structural depolarization up to a certain 
limit until $\tau_f$ becomes too large.  For $\Delta\phi > 90^\circ$ the internal 
rotation initially increases the depolarization as $\tau_f$ increases but 
later can also go into a region of 'inverse depolarization'.  

Figure 1(b) shows that for simple field geometries, such as the two 
cell system shown here, the observed EVPA for internal rotation is linear 
in $\lambda^2$ with $\chi = \chi_0 + \tau_f/4$ with additional $90^\circ$ flips 
when the polarized flux passes through zero\footnote{Note that for purely 
{\em external} rotation in a uniform screen $\tau_f$ is also 
proportional to $\lambda^2$, but $\chi = \chi_0 + \tau_f/2$.  However, that is
not the case we consider in this paper.} \citep[e.g.][]{B66}.  More 
complicated geometries which include a range of $\tau_f$ values along different 
lines of sight are expected to diverge from a $\lambda^2$ law after total rotations 
$\gtrsim 45^\circ$ \citep{B66}, although the total rotation at which the
non-$\lambda^2$ behavior becomes apparent will vary from case to case (see
the lower panels of Figure 2 as examples).

\subsection{Inverse Depolarization in Helical Fields}
To evaluate the efficiency of this mechanism in a purely helical field
geometry, we numerically solved the full equations of radiative
transfer \citep{JOD77,Jones88} for a variety of combinations of 
toroidal ($f_t$) and uniform ($f_u$) field components in an optically
thin cylinder.  The uniform
field is poloidally directed along the jet axis and has constant magnitude 
throughout the cylindrical cross-section of the jet.  The magnitude of
the toroidal field component grows linearly with distance from the jet axis from 
0 to $f_t$ at the outside edge of the cylinder, 
consistent with field generated by a current which is carried uniformly in
the jet volume.  A more detailed description of the radiative transfer 
simulation and magnetic field parameters is given by \citet{HL09}.  The
simulated jets are relativistic, with bulk Lorentz factor $\Gamma = 10$,
and viewed at one-half the optimum angle for superluminal motion 
$\theta = 1/2\Gamma$.  This intermediate viewing angle is the most
probable in a flux-density limited sample \citep{LM97}; however, we
note that a wide range of viewing angles is possible and may
strengthen or weaken the inverse depolarization effect in various
cases.  Internal Faraday depth is provided by the low end of the
relativistic particle spectrum\footnote{We note that 
the same results would be obtained by introducing thermal particles to
do the rotation, as it is only the amount of rotation 
along each line of sight that determines the emergent polarization.},  
resulting in $\langle\tau_f\rangle \sim 2\lambda^2$, averaged
over all lines of sight, for the simulations shown in Figure 2. 

Figure 2 plots the integrated polarization
from the entire jet cross-section as a function of $\lambda^2$ for a 
variety of combinations of $f_u$ and $f_t$.
Even in these integrated results, the inverse depolarization 
effect is clearly seen, for at least some range of wavelengths,
in half of the cases shown.  Note that there is not necessarily
any correlation in the integrated results between the amount of
inverse depolarization and the observed Faraday rotation (given
in the lower panel of Figure 2).  This lack of correlation in
the integrated results is primarily due to variation in both
Faraday depth and field order across the jet.   

In a helical magnetic field, the offset angle, $\Delta\phi$,
between the back and front of the jet varies with location
along the cylindrical cross-section.  Additionally, the
internal Faraday depth, $\tau_f$, is produced by the 
line-of-sight components of {\em both} the poloidal (uniform)  
and toroidal fields.  The contribution to $\tau_f$ from the 
toroidal field will also vary with location along the cylindrical 
cross-section, changing the sign of its contribution at 
the mid-point of the jet.  The details of how these factors
combine depend on a variety of factors, but the important
point is that in a helical magnetic field, the inverse
depolarization effect can vary strongly as a function of location
along the jet cross-section.  Detecting and modeling this 
variation requires sensitive observations at several 
wavelengths; however, in Figure 3 we demonstrate that 
the effects could be observable even with only two beam-widths
of resolutions across the jet.  

Figure 3 shows a partially resolved, cylindrical jet 
cross-section in relative total intensity (solid curve) 
and fractional polarization at a range of wavelengths (symbols).  
Even with only two beam-widths of resolution, the inverse
depolarization effect clearly varies with location along 
the jet, being extremely strong at the bottom of the jet (left
of figure) and turning to normal depolarization at the
top of the jet (right of figure).  This is a
particular case, and, as noted above, the details will vary
depending on the combination of poloidal and toroidal fields, jet viewing angle,
and average Faraday depth.  It is interesting to note 
that a jet viewed perpendicular to its axis in the fluid frame, $\theta=1/\Gamma$,
can still have a significant inverse depolarization effect, even though
it has no component of poloidal field along the line of sight.  This
can happen if the field crossing angle is $\Delta\phi \simeq \pm90^\circ$ above and below
the jet axis.  Although the $\tau_f$ provided by
the toroidal field reverses across the axis, either sign of $\tau_f$ 
will increase the net polarization from near zero giving inverse 
depolarization on both the top and bottom of the jet.

\subsection{Inverse Depolarization in Random Fields}

The structural inhomogeneities that are necessary for 
inverse depolarization can also be produced
stochastically in a randomly tangled magnetic field. 
The effect should be strongest in those cases where structural
cancellation leads to low net polarization so that internal
rotation can effectively increase the fraction of polarization
with wavelength.  We have simulated this scenario with a
$N\times N\times N$ cube of cells, each with a randomly oriented
magnetic field.  The results for $N=5$ and $N=15$ are plotted
in Figure 4 where we show the percentage of cases with significant
inverse depolarization between two wavelengths (top panel) and the
average polarization of those cases showing the inverse 
depolarization (bottom panel). Significant inverse depolarization
occurs for $20-30$\% of cases for both $N=5$ and $N=15$, with and
without a modest ``shock'' applied to the entire cube (see caption), 
for a range of mean Faraday depths.  However, the mean fractional
polarization in the $N=15$ cases with inverse depolarization is
very low, $\simeq 1-2$\%.  The $N=5$ cases are larger, 
$\simeq 5-8$\% polarization, and more consistent with the 
levels of polarization seen in the jets showing this effect,
although levels up to $20$\% are observed \citep{HL12}. 

\subsection{Internal Faraday Depth}
\label{s:tau_f}

Internal to the jet, Faraday depth may be due to either a population 
of ``cold'' thermal electrons with number density, $n_c$, or the low-energy 
end of the relativistic particle distribution with number density,
$n_r$.  The relative contributions to the Faraday depth from the cold and 
relativistic particles are given by \citet{JOD77}:

\begin{equation}
\frac{\tau_f^{(r)}}{\tau_f^{(c)}} = 2\alpha\frac{\alpha+3/2}{\alpha+1}\frac{\ln\gamma_i}{\gamma_i^2}
\left(\frac{n_r}{n_c}\right)
\end{equation}

\noindent where $\alpha$ is the optically thin spectral index defined by $S\propto\nu^{-\alpha}$.  
The relativistic contribution assumes a power-law distribution of particles with 
$dn_r = K\gamma^{-(2\alpha+1)}d\gamma$ with a lower cutoff $\gamma_i$.  For cold particles
alone, we have the following familiar expression (in cgs units) \citep[e.g.][]{RL79}.

\begin{equation}
\tau_f^{(c)} = \frac{e^3}{\pi m^2c^4}\frac{\delta^2\lambda^2}{(1+z)^2}\int n_c\mathbf{B\cdot dl}
\end{equation}  

Note that $\lambda$ in the above expression is the {\em observed} wavelength.  
Internal to the jet, the rotating plasma sees a much longer wavelength because 
the jet emission is Doppler boosted in the observer's frame by factor, $\delta/(1+z)$.
It is interesting to note that, for typical blazars with $\delta \gtrsim 10$, 
low-energy particles internal to the jet (whether cold or relativistic) 
can be hundreds of times more effective at Faraday rotation than rotating plasma 
external to the jet which is not moving relativistically in relation to the observer
(assuming equivalent field strengths and path lengths).

The inverse depolarization effect proposed here requires only modest Faraday 
depths of order $\tau_f \simeq$ a few, and indeed large Faraday depths would 
effectively depolarize jets, which is not observed in blazars.  Suppression 
of large internal Faraday depths requires some combination of the following conditions:
(1) low number density of the rotating particles, both thermal {\em and} low-energy
relativistic, (2) pairing of low-energy relativistic electrons with low-energy 
relativistic positrons without a significant thermal component, or (3) a large 
number of magnetic field reversals along the line of sight. Study of this 
effect which implies modest but not large internal Faraday rotation, and provides 
constraints on the magnetic field order, may yield new information on the 
low-energy particle population of relativistic jets.

\section{Summary and Conclusions}

We have shown that observations of inverse depolarization in optically
thin jet features \citep{HL12,H02}, can 
potentially be explained by a simple physical model which combines
structural inhomogeneities in the magnetic field along the
line of sight with internal Faraday rotation. Structural inhomogeneities
of the right type are naturally produced by helical magnetic fields.
Random magnetic fields can also produce this effect, but must be tangled
on long length scales in the jet to be consistent with the modest to high 
levels of fractional polarization observed in these features. Real jets
likely contain some combination of randomly tangled and ordered field 
components \citep[e.g.][]{Hughes05,LCB06}; however, it is intriguing that 3 of 
the 4 jets which appear to show this effect also have transverse rotation
measure gradients, a potential signature of helical magnetic fields 
\citep{A02,ZT05,G08,HL12}, although for 3C\,120 the evidence 
favors an interaction with the external medium as the source of the
rotation measure gradient \citep{G08}.
In the case of inverse depolarization produced by helical magnetic fields, we have 
demonstrated that the transverse structure of inverse depolarization could be 
measured even in partially resolved jets, and the recently expanded bandwidth of
the Very Long Baseline Array may allow highly sensitive, multi-wavelength
observations of transverse jet profiles to test these predictions in
the near future.  

\acknowledgements

The author would like to thank John Wardle, Talvikki Hovatta, Matt Lister,
Margo Aller and the other members of the MOJAVE collaboration for helpful 
discussions. This work has been supported by NSF grant AST-0707693


%
%

\begin{figure}[onecolumn]
\includegraphics[scale=0.65,angle=-90]{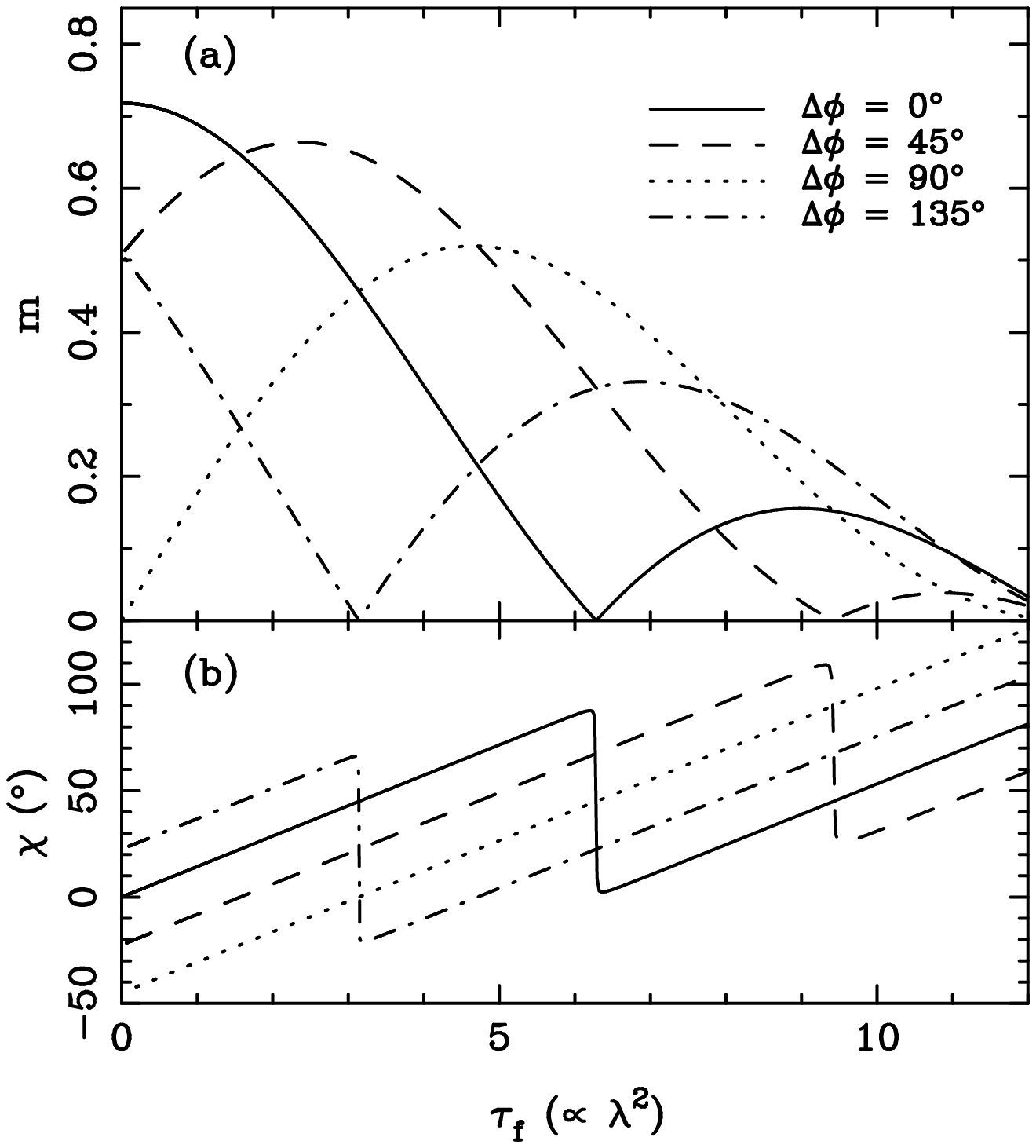}
\figcaption{\label{f:2cell}
Plot of fractional linear polarization (top panel) and EVPA (bottom panel) as a function of Faraday 
depth, $\tau_f \propto \lambda^2$, for a simple two-cell system along the line of sight.  Each line shows
the emergent polarization for a different offset angle, $\Delta\phi$, for the projected magnetic 
field between the far and near cells.}
\end{figure}

\begin{figure}
\includegraphics[scale=0.65,angle=-90]{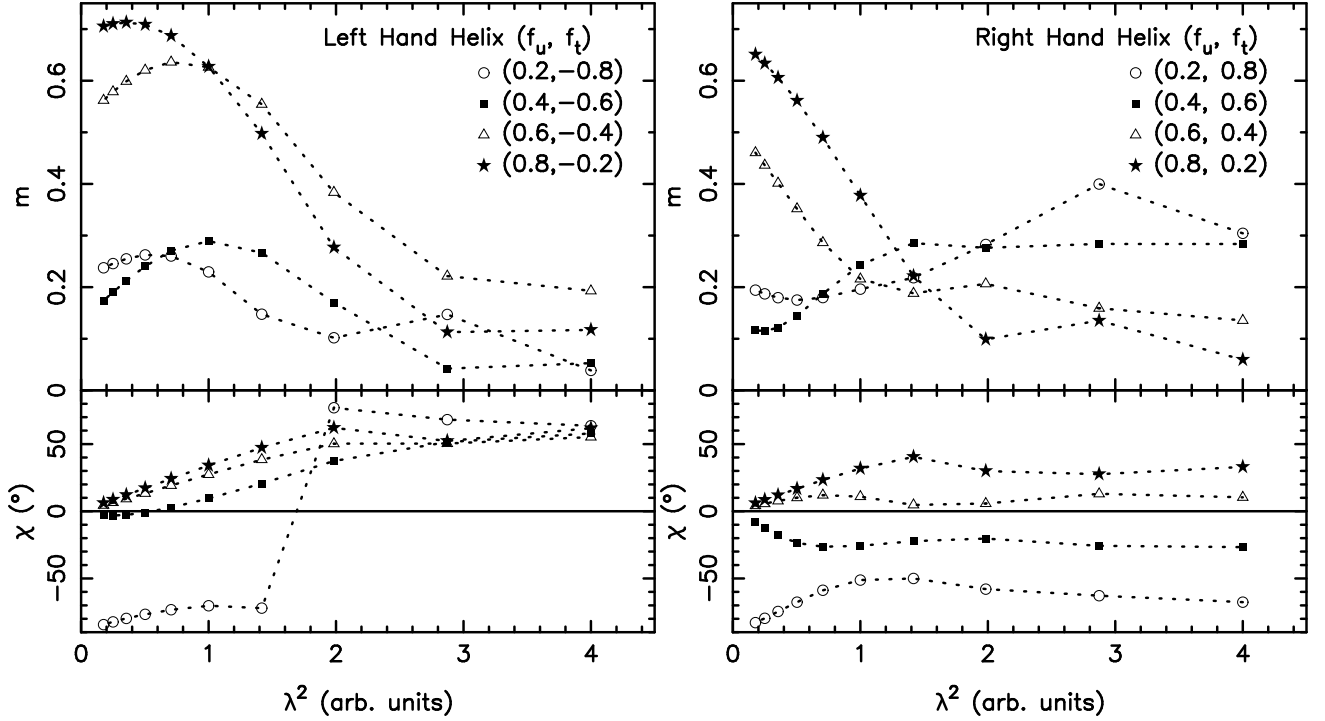}
\figcaption{\label{f:net_helix}
Integrated polarization as a function of $\lambda^2$ for cylindrical jets with pure helical 
magnetic fields.  Jets with various combinations of uniform magnetic field, along the jet 
axis, $f_u$, and toroidal magnetic field, $f_t$ are plotted.  All of the jets here are identical
except for their $(f_u,f_t)$ combination, and all are at an angle $\theta = 1/2\Gamma$ to the observers
line of sight, inside the optimum angle for superluminal motion.}
\end{figure}

\begin{figure}
\includegraphics[scale=0.5,angle=0]{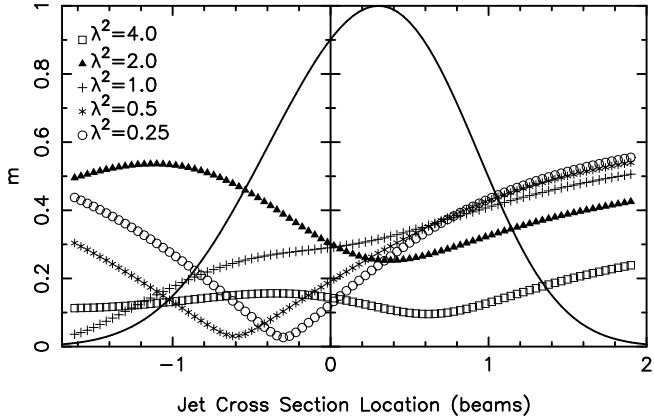}
\figcaption{\label{f:helix}
Fractional polarization, $m$, as a function of transverse jet position for a partially resolved jet 
where the jet diameter is twice the FWHM of the observing beam.  The $(f_u,f_t) = (0.4,-0.6)$
case from figure \ref{f:net_helix} is plotted for several values of $\lambda^2$.  The single solid
line represents the Stokes $I$ intensity normalized to a maximum of 1.0.  Note that the asymmetric
profile in Stokes $I$ is similar to that predicted by \citet{CB11} for a helical magnetic 
field viewed at $\theta=1/2\Gamma$.
}
\end{figure}

\begin{figure}
\includegraphics[scale=0.65,angle=-90]{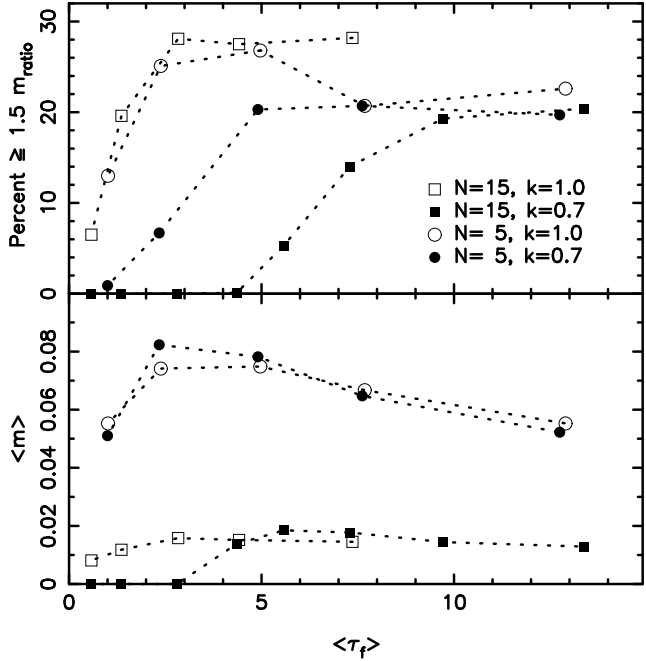}
\figcaption{\label{f:random}
Top panel shows the percentage of 1000 runs where the fractional polarization ratio, $m_{ratio}$ 
between $\lambda^2=2.0$ and 
$\lambda^2 = 1.0$ is $\geq 1.5$ indicating significant inverse depolarization.   
Integrated results from cubes of random field of 
dimensions $5\times5\times5$ and $15\times15\times15$ are plotted with and without a moderate shock 
where unit length has been shortened to length $k$ along the jet axis \citep{WCRB94}.  
Each case is plotted for a variety of mean Faraday depths, $<\tau_f>$.  
The bottom panel shows the mean fractional polarization at the longest wavelength 
for the runs where $m_{ratio} \geq 1.5$. }
\end{figure}

\end{document}